\newif\ifarxiv
\arxivfalse 
\ifarxiv
  \PassOptionsToClass{nonacm}{acmart}
\fi

\documentclass[sigconf, balance=false]{acmart}

\usepackage{xltabular}
\usepackage[table]{xcolor} 
\usepackage{array}
\usepackage{listings}
\AtBeginDocument{%
  \providecommand\BibTeX{{%
    \normalfont B\kern-0.5em{\scshape i\kern-0.25em b}\kern-0.8em\TeX}}}

\definecolor{brown}{rgb}{0.59, 0.29, 0.0}
\definecolor{darkgray}{rgb}{0.59, 0.59, 0.59}
\definecolor{tablegray}{gray}{.9}

\definecolor{nicecolor}{rgb}{0, 0.58, 0.51}

\usepackage{algorithm}
\usepackage{algpseudocode}
\usepackage{amsmath}

\algblockdefx[Inputs]{Inputs}{EndInputs}
  {\textbf{Inputs:}}
  {} 

\algblockdefx[Outputs]{Outputs}{EndOutputs}
  {\textbf{Outputs:}}
  {} 

\def\signed #1{{\leavevmode\unskip\nobreak\hfil\penalty50\hskip2em
  \hbox{}\nobreak\hfil(#1)%
  \parfillskip=0pt \finalhyphendemerits=0 \endgraf}}

\newsavebox\mybox

\newcommand\highlight[1]{\textcolor{red}}

\usepackage[utf8]{inputenc}
\usepackage{diagbox}
\usepackage{colortbl}

\usepackage{tabularx}
\usepackage{multirow}

\usepackage{color}
\usepackage{soul}

\usepackage{xspace}
\usepackage{enumitem}
\usepackage{mathtools}
\usepackage{commath}

\usepackage{amssymb}
\usepackage{pifont}

\newcommand{\customtilde}{{\raise.17ex\hbox{$\scriptstyle\sim$}}}

\newcommand{\etal}{et~al.\xspace}
\newcommand{\eg}{\textit{e.g.},\xspace}
\newcommand{\ie}{\textit{i.e.},\xspace}

\usepackage{xparse}

\newcommand{\system}{Dramarrator}

\definecolor{speechcolor}{HTML}{4c78a8}
\definecolor{audiotagcolorold}{HTML}{72b7b2}
\definecolor{audiotagcolor}{HTML}{44706D}
\definecolor{audiotagbg}{HTML}{E8F5F4}
\definecolor{scenecolor}{HTML}{e45756}
\definecolor{scenebg}{HTML}{FDECEA}
\definecolor{sfxcolor}{HTML}{f58518}
\definecolor{sfxbg}{HTML}{FEF5E8}
\definecolor{musiccolor}{HTML}{54a24b}
\definecolor{musicbg}{HTML}{EEF7EC}

\newcommand{\audiotag}[1]{%
  {\setlength{\fboxsep}{1pt}\colorbox{audiotagbg}{\textcolor{audiotagcolor}{\texttt{[#1]}}}}}

\definecolor{deletioncolor}{HTML}{eaecec}
\definecolor{insertioncolor}{HTML}{f6e5f1}
\definecolor{fillerwordcolor}{HTML}{dcefdc}
\definecolor{repetitioncolor}{HTML}{e1efc4}
\definecolor{emphasiscolor}{HTML}{faefb7}
\definecolor{clarificationcolor}{HTML}{f8ce88}
\definecolor{informationcolor}{HTML}{f9a77c}
\definecolor{deletedgray}{HTML}{7F7F7F}

\usepackage{soulpos}

\ulposdef{\hlst}{%
  \rlap{\textcolor{deletioncolor}{\rule[-.75ex]{\ulwidth}{2.5ex}}}
  \rule[.45ex]{\ulwidth}{.1ex}
}

\soulregister{\color}{1}

\definecolor{emphasislight}{HTML}{fcf6d2}  
\definecolor{emphasismedium}{HTML}{faefb7} 
\definecolor{emphasisstrong}{HTML}{fce45a} 

\newcommand{\alphaval}[2]{{\small $p\,#1\,#2$}}

\ifarxiv
  \setcopyright{none}
  \renewcommand\footnotetextcopyrightpermission[1]{}
\else
    \copyrightyear{2026}
    \acmYear{2026}
    \setcopyright{cc}
    \setcctype{by-nc-nd}
    \acmConference[UIST '26]{The 39th Annual ACM Symposium on User Interface Software and Technology}{November 02--05, 2026}{Detroit, MI, USA}
    \acmBooktitle{The 39th Annual ACM Symposium on User Interface Software and Technology (UIST '26), November 02--05, 2026, Detroit, MI, USA}
    \acmDOI{10.1145/3830398.3830546}
    \acmISBN{979-8-4007-2856-3/2026/11}

    \makeatletter
    \def\@ACM@copyright@check@cc{}
    \makeatother
\fi

\begin{document}

\title{\system: Object-Based Audio Editing for Audio Drama Production from Books}


\author{Karim Benharrak}
\authornote{Work done during an internship at Adobe Research.}
\orcid{0009-0002-3279-5664}
\email{karimbenharrak@berkeley.edu}
\affiliation{%
  \institution{University of California, Berkeley}
  \city{Berkeley}
  \state{CA}
  \country{USA}
}

\author{Oriol Nieto}
\orcid{0000-0001-6459-7609}
\email{onieto@adobe.com}
\affiliation{%
  \institution{Adobe Research}
  \city{San Francisco}
  \state{CA}
  \country{USA}
}

\author{Bryan Wang}
\orcid{0000-0001-9016-038X}
\email{bryanw@adobe.com}
\affiliation{%
  \institution{Adobe Research}
  \city{Seattle}
  \state{WA}
  \country{USA}
}

\author{Zeyu Jin}
\orcid{0000-0003-0161-5915}
\email{zejin@adobe.com}
\affiliation{%
  \institution{Adobe Research}
  \city{San Francisco}
  \state{CA}
  \country{USA}
}

\author{Amy Pavel}
\orcid{0000-0002-3908-4366}
\email{amypavel@eecs.berkeley.edu}
\affiliation{%
  \institution{University of California, Berkeley}
  \city{Berkeley}
  \state{CA}
  \country{USA}
}

\renewcommand{\shortauthors}{Benharrak, et al.}

\begin{abstract}
Audio dramas weave dialogue, sound effects, and music into immersive stories. Creators often adapt books into audio dramas, but this process remains labor-intensive, requiring them to interpret source material, author scripts, generate audio assets, and assemble them on a timeline. Because story elements like characters and scenes manifest across many interdependent assets, a single change can ripple into manual updates across the entire project. We present \system{}, an audio drama authoring tool built around object-based audio editing, where these story elements are represented as editable objects. \system{} extracts these objects from a book, generates linked audio assets (speech, sound effects, and music), and composes a multi-track audio drama. Edits to any object (\eg a character's voice) automatically propagate to all dependent assets. In a user study with professionals (N=8), \system{} significantly lowered task load when creating audio dramas. A listener study (N=300) shows that creator-refined output from \system{} approaches the quality of productions made with existing professional tools, and an exploratory study (N=3) suggests object-based editing lowers entry barriers and generalizes beyond audio dramas.

\end{abstract}



\begin{teaserfigure}
  \includegraphics[width=\textwidth]{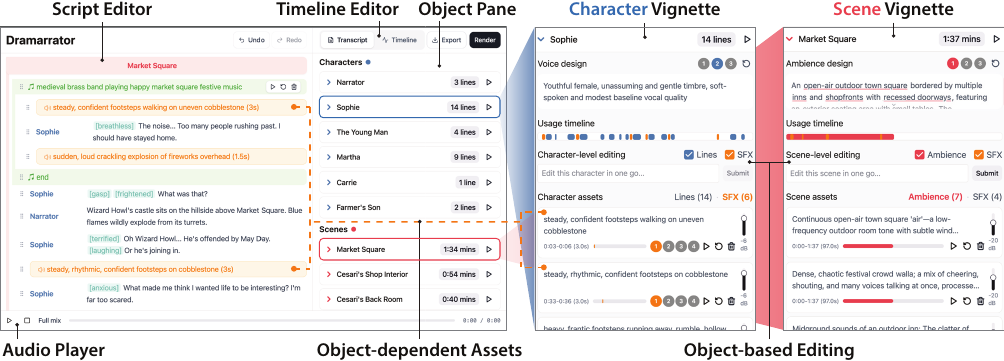}
  \caption{\system{}'s interface for object-based audio drama authoring. The Script Editor (left) represents a multi-track audio timeline via a transcript with dialogue, sound effects, and music, nested within scenes. The Object Pane (right) visualizes all \textcolor{speechcolor}{Character} and \textcolor{scenecolor}{Scene} objects: a Character Vignette exposes a character's voice design and their linked speech lines and SFX, and a Scene Vignette exposes a scene's ambience design and its linked ambience and SFX assets. When creators edit an object (\eg changing a character's voice), \system{} automatically propagates edits across all affected object-dependent assets.}
  \label{fig:teaser}
\end{teaserfigure}


\maketitle

\section{Introduction} Audio dramas adapt source material (\eg books) into immersive experiences told through dialogue, sound effects (SFX), and music~\cite{wayland2020bombs, rattigan_theatre_2002, crook_radio_2002}. Adapting books into audio dramas requires creators to decompose expository prose into dialogue and continually decide how to convey actions, environments, and atmospheres through speech, sound effects, and music~\cite{toscan2023writingaudiodrama, wayland2020bombs}. For example, a spoken line ``\textit{I'm opening this crate}'' may be replaced with a sound effect that conveys the same action. Recent generative AI advances dramatically lowered the cost of producing individual audio assets from text prompts~\cite{elevenlabsv3, borsos_audiolm_2023, elevenlabs, suno, adobefirefly}. However, assembling assets into a coherent narrative and iteratively refining it remains challenging.

Current audio editing tools such as Adobe Audition~\cite{adobeaudition}, Pro Tools~\cite{protools}, and Logic Pro~\cite{logicpro} represent audio dramas as numerous isolated audio clips on a linear timeline, with limited explicit connection to the story constructs (\eg characters, scenes, story arcs) that creators reason about~\cite{wayland2020bombs, de_fossard_writing_2005}. For example, a character ``Tom'' may span many speech clips across multiple scenes, yet existing tools do not model ``Tom'' as a coherent entity. Thus, single high-level decisions, such as changing a scene's weather or rewriting a character's dialogue tone, require manually locating, regenerating, and replacing every affected clip across all timeline tracks, which discourages revision of early decisions and raises the barrier to multi-episode production.

Prior work has shown that editing through higher-level abstractions simplifies workflows in creativity support tools~\cite{li_beyond_2023, beaudouin-lafon_instrumental_2000} by grouping lower-level elements into editable objects across programming~\cite{rentsch_object_1982, stefik_object-oriented_1985}, design~\cite{xia_object-oriented_2016, figma, xia_collection_2017}, and multimedia editing~\cite{kim_cells_2023, xia_crosspower_2020, han_passages_2022, yeh_craftml_2018}. These abstractions, however, typically compose objects from elements that share a uniform representation (\eg code instances, design components, or text passages). In audio dramas, narrative constructs such as characters and scenes comprise heterogeneous asset types (\eg a character can be expressed through their script, vocal delivery, and accompanying sound effects). A single narrative change, such as altering a character's personality, must therefore be translated into distinct edits across these asset types.

To achieve this, we introduce \textbf{object-based audio editing} and present \system{}, an end-to-end pipeline and authoring interface for adapting books into audio dramas. \system{} models narrative constructs as editable \textit{objects}: a Character object such as ``Tom'' has attributes like voice design, and a Scene object such as ``Market Square'' has attributes like ambience design. Each generated audio asset on the timeline (\eg a speech clip, a sound effect) is linked back to the object it belongs to. This structure enables editing at the narrative level: changing Tom's voice design to be raspier, for example, automatically regenerates all of Tom's speech clips across the entire project. \system{}'s pipeline extracts these objects from the book's source text, adapts the text into an audio drama script, and orchestrates multiple generative models to produce all audio assets. Creators can then review and refine both objects and individual assets through \system{}'s interface (Figure~\ref{fig:teaser}).

In a user study with 8 professional audio drama producers, \system{} significantly lowered the required task load when creating audio dramas compared to existing tools.
Professionals attributed time savings to object-based editing: a single object change (\eg updating a character's voice) propagates automatically to all dependent assets, replacing up to 30 manual edits at once.
Professionals also reported higher creative exploration, since \system{} lets them iteratively shape characters and scenes while hearing them in the final audio drama, whereas their existing sequential workflows require all character and scene decisions to be fixed early on in their process.
A listener study ($N=300$) shows that audio dramas refined by professionals in \system{} are comparable to those produced with existing tools on engagement, audio and story quality, and character and scene consistency.
An exploratory study involving 3 novice creators from adjacent domains (game, RPG, and puzzle design) shows that \system{} enables novices without audio-editing expertise to create audio dramas. Their feedback also suggests that object-based editing may generalize to other creative domains organized around narrative constructs such as characters and scenes.

In summary, we contribute:
\begin{itemize}
    \item Object-based audio editing, a concept that links narrative objects (\eg characters, scenes) with editable attributes (\eg voice, ambience) and object-dependent assets (\eg speech, sound effects, music) to enable high-level edits that automatically propagate across the entire audio.
    \item \system{}, an audio drama authoring system that generates speech, music, and sound effects from a book, automatically assembles them into an audio drama, and allows creators to steer the output through object-based editing.
    \item A user study (N=8) showing that \system{} significantly lowers task load when creating audio dramas compared to professionals' existing tools, complemented by a listener study (N=300) evaluating output quality and an exploratory study (N=3) probing novice perceptions and the generalizability of object-based editing to other storytelling domains.

\end{itemize}

\section{Related Work}
We build upon prior work on audio editing tools, editing with higher-level abstractions, and automated audio generation.

\subsection{Audio Editing Tools} Digital audio workstations (DAWs) such as Pro Tools~\cite{protools}, Logic Pro~\cite{logicpro}, and Adobe Audition~\cite{adobeaudition} support multi-track audio editing by organizing speech, sound effects, and music across a linear timeline. While powerful, these tools impose steep learning curves and demand considerable manual effort to navigate and assemble numerous individual audio clips. To mitigate some of these challenges, especially for spoken content, text-based audio editing interfaces~\cite{descript, benharrak_talkless_2025, rubin_content-based_2013, shin_dynamic_2016} enable users to modify content by editing a script that is time-aligned to the corresponding speech. For example, Rubin~\etal~\cite{rubin_content-based_2013} presented one of the earliest content-based editing tools for audio story editing, and Shin~\etal~\cite{shin_dynamic_2016} bidirectionally linked scripts to audio so that edits in one representation are reflected in the other. This progression from waveform to text-based editing reflects a broader shift toward more semantic, narrative-level representations of audio. Our work features both existing editing paradigms while further advancing the trajectory toward object-based editing, where users manipulate narrative constructs, such as characters and scenes, rather than individual audio clips or words in transcripts.

\subsection{Editing via Higher-Level Abstractions}
In creative projects, a single conceptual change often requires updating many dependent elements. Changing a brand color in a design system or a character's personality in a story means manually finding and modifying every affected artifact. Prior work facilitates this by grouping related elements into higher-level objects that propagate edits automatically~\cite{li_beyond_2023, beaudouin-lafon_instrumental_2000}, as seen in programming~\cite{rentsch_object_1982, stefik_object-oriented_1985}, text~\cite{kim_cells_2023, han_passages_2022}, visual~\cite{xia_object-oriented_2016, xia_collection_2017, suh_storyensemble_2025}, and 3D tools~\cite{yeh_craftml_2018}. Modifying a class in object-oriented programming updates all its instances; editing a component in Figma~\cite{figma} propagates changes to every screen that uses it. These abstractions, however, typically group elements that share a uniform representation (\eg code instances, design components). In audio dramas, a single narrative construct, such as a character's personality, is realized through heterogeneous assets spanning text (dialogue) and audio (voice style, SFX, music). Editing at this level requires knowing not only \emph{which} assets to change but \emph{how} to transform a narrative-level change into appropriate modifications across modalities. To this end, \system{} introduces \textit{object-based audio editing}. \textit{Objects} represent narrative constructs (\eg characters, scenes) with editable \textit{attributes} (\eg voice or ambience design) and \textit{object-dependent assets} (\eg speech, sound effects, music). \system{} also gives users freedom to navigate the ladder of abstraction~\cite{li_beyond_2023}: high-level object changes automatically propagate to all dependent assets, while low-level edits to individual assets can be made via a transcript or timeline editor.

\subsection{Automated Audio Generation}
Recent advances in generative AI have made audio production faster and more accessible by using text prompts to generate speech~\cite{elevenlabsv3, laban2022newspod, google_notebooklm}, sound effects~\cite{borsos_audiolm_2023, kreuk_audiogen_2023}, or music~\cite{copet_simple_2023, agostinelli_musiclm_2023}. However, these models typically generate individual audio clips rather than structured, multi-track compositions. Recent work orchestrates multiple models to combine speech, sound effects, and music from text instructions~\cite{liu_wavjourney_2025, guo_audiostory_2025, xu_mm-storyagent_2025}, but these systems are limited to short outputs of approximately 2--3 minutes and produce a single flattened audio file, preventing creators from editing or rearranging individual elements after generation. Most closely related to our work, SoundStager~\cite{yoo_soundstager_2026} generates editable layered soundscapes but requires video input where dialogue and visuals already establish narrative context. In audio dramas, by contrast, the auditory channel is the only medium: speech, sound effects, and music must jointly convey the narrative, and characters and scenes must remain consistent across productions spanning tens of minutes to hours. To address these requirements, \system{} employs object-based editing to orchestrate multiple generative models into long-form, multi-track audio dramas, while exposing the editable structure for creators to fine-tune results at both the narrative and asset levels.

\section{Design Formulation}
To understand existing practices and challenges of creating audio dramas, we analyzed literature from industry professionals~\cite{wayland2020bombs, toscan2023writingaudiodrama, bernaerts_audionarratology_2021, de_fossard_writing_2005, crook_radio_2002, rattigan_theatre_2002}, expert interview videos on YouTube~\cite{yt-bbcscriptwriting, yt-boothjunkieaudiodramasounddesign, yt-bbcwritingforradio, yt-cactusaudiodrama, yt-foolwritingforaudiofiction, yt-kennethbranagh, yt-radioushowtowrite, yt-ryanaudiodrama, yt-writingcomics}, and online community discussions\footnote{www.reddit.com/r/audiodrama/}.
We analyzed all insights using thematic analysis~\cite{braun_using_2006, braun_reflecting_2019, mayring_qualitative_2021}. 
We first extracted key insights, then moved back and forth through the material using axial coding principles~\cite{corbin_basics_2014}, and finally iteratively clustered, split, and merged codes.
We collected best practices (Table~\ref{tab:formative_results}) and derived 4 design goals (\textbf{DG1}-\textbf{DG4}) for a system that supports creating audio dramas from books:
\medskip

\noindent \textbf{DG1: Adapt book text into scripts that follow audio drama conventions.}
Books contain internal monologue, descriptive narration, and extended exposition that cannot be conveyed through sound and bore listeners if narrated verbatim~\cite{rattigan_theatre_2002, bernaerts_audionarratology_2021, toscan2023writingaudiodrama, wayland2020bombs}.
Thus, adapting a book into an audio drama requires restructuring the source text so that internal thoughts are externalized into dialogue or sound events, dialogue is short and natural, exposition delivered just-in-time, and narration that restates what sound already conveys is removed~\cite{toscan2023writingaudiodrama, wayland2020bombs, crook_radio_2002}.
We aim to support creators by automatically producing scripts that follow existing conventions.
\medskip

\noindent \textbf{DG2: Create distinct and consistent characters and scenes.}
Listeners identify characters and locations entirely through sound.
Because inconsistency across characters or scenes along the story breaks immersion~\cite{de_fossard_writing_2005, crook_radio_2002}, each character needs a distinctive voice (\eg pitch, pace, accent, word choice) and associated sound effects (\eg footsteps)~\cite{wayland2020bombs, crook_radio_2002, de_fossard_writing_2005, rattigan_theatre_2002}.
Similarly, each scene needs a distinctive ambience soundscape (\eg rain, traffic), sound effects for character actions (\eg footsteps, a slamming door), and music that matches the scene's mood~\cite{wayland2020bombs, toscan2023writingaudiodrama, crook_radio_2002}.
We aim to support creators in defining characters and scenes as persistent narrative constructs that remain consistent across the full audio drama.
\medskip

\noindent \textbf{DG3: Produce a balanced mix with intelligible dialogue.}
Because audio dramas consist of multiple tracks mixed together, a challenge is to ensure that listener attention remains on the dialogue, which should be easy to follow, without distracting sound elements~\cite{wayland2020bombs, rattigan_theatre_2002}.
Professionals follow a mixing hierarchy in which they place the volume of tracks relative to the dialogue (\ie loudest).
For example, sound effects are placed slightly below the volume of dialogue, music further below, and ambience at the lowest level~\cite{de_fossard_writing_2005, crook_radio_2002, wayland2020bombs}, with ambience and music reducing its volume when characters speak~\cite{wayland2020bombs, toscan2023writingaudiodrama}.
We aim to produce a multi-track audio drama mix that makes sure dialogue is intelligible without abrupt sounds or transitions.
\medskip

\noindent \textbf{DG4: Support editing of narrative constructs at different production stages.}
Audio drama production traditionally follows a sequential workflow that moves from high-level to low-level decisions.
Creators first make high-level narrative decisions such as writing the script, designing the characters and scene, and then move to lower-level tasks such as audio production, timeline arrangement, and mixing \cite{crook_radio_2002, wayland2020bombs}.
Higher-level changes in later production stages are tedious and time-consuming as they may require substantial changes as all dependent audio assets must be manually identified, re-produced, and replaced across the entire project.
We aim to support creators in revising narrative constructs such as characters and scenes at any time without having to restart from an earlier production stage.

\section{\system}
Based on our design goals (\textbf{DG1}-\textbf{DG4}), we developed \system{}, which introduces object-based audio editing where narrative constructs (\eg characters, scenes) are represented as objects with editable attributes (\eg voice, ambience) linked to their dependent audio assets.
We introduce object-based audio editing in (1) an editing interface where creators review and refine objects to propagate high-level changes across the project, and (2) in an automatic pipeline that extracts objects from book text to generate multi-track audio drama adaptations.

\subsection{Object-Based Audio Editing}
\label{sec:object-based}
\system{} introduces object-based audio editing based on three concepts: \textbf{objects}, \textbf{attributes}, and \textbf{object-dependent assets}.
Objects represent narrative constructs such as characters and scenes.
Each object holds attributes that describe properties of that object, for example a character's voice design or a scene's ambience design.
Finally, each object contains a set of object-dependent assets that keep track of isolated audio assets on the timeline (\eg speech, sound effects, music) that depend upon the object.
Every generated audio asset on the timeline is linked to one or more objects at creation time.
A character's dialogue lines and sound effects (\eg their footsteps) are linked to that character object.
All sound effects, ambience layers, and music cues that occur while a scene is active are linked to that scene object.
A sound effect can be linked to both a character and a scene at the same time.
For example, Martha's footsteps in the Market Square scene are linked to both the Martha character object and the Market Square scene object. 

\begin{figure}
    \centering
    \includegraphics[width=\linewidth]{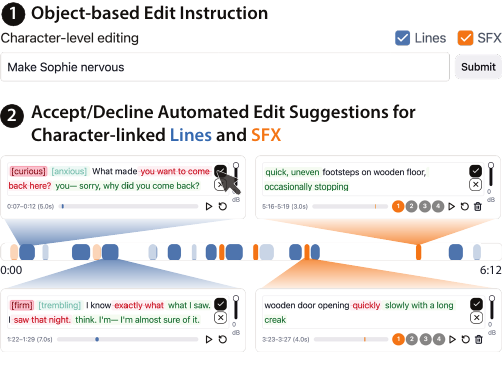}
    \caption{\system's object-based editing automatically propagates edits on character or scene objects across linked assets that require updates. Specifying a higher-level edit instruction to modify a character (1), triggers \system{} to find and suggest edits across all affected assets on the timeline that users can choose to accept or reject (2).}
    \label{fig:object-based-editing}
\end{figure}

When a creator modifies an object's attribute (\eg~\textit{``make Sophie nervous''}), \system{} uses an LLM to identify all dependent assets that require updates and proposes modification suggestions (\eg rewriting firm spoken lines to sound nervous) for creators to accept or decline (Figure~\ref{fig:object-based-editing}). Rather than manually locating and updating all affected audio assets, creators make a single object change that propagates automatically across all object-dependent assets (\textbf{DG4}).

\subsection{\system{} Interface}
\system{}'s interface consists of three components: an \textbf{Object Pane} for reviewing and editing character and scene objects, a \textbf{Script Editor} for script-based editing, and a \textbf{Timeline Editor} for timeline-based editing (Figure~\ref{fig:teaser}).
We follow Leila, who creates and edits audio dramas using \system{}.
\medskip

\noindent The \textbf{Object Pane} displays all extracted \textcolor{speechcolor}{Character} and \textcolor{scenecolor}{Scene} objects as editable vignettes with their attributes and object-dependent assets for quick review and editing (\textbf{DG2, DG4}).
Character vignettes show a voice design attribute alongside three generated voice previews, and all object-dependent assets are accessible via a \textit{Lines} and \textit{SFX} tab.
Scene vignettes show an ambience design attribute as an annotated prompt where individual words are linked to their corresponding ambience layers within the \textit{Ambience} tab to make the prompt-to-sound mapping explicit. An \textit{SFX} tab lists all sound effects in that scene.
Both vignettes support natural-language edit instructions (\eg \textit{``make Sophie nervous''}), upon which \system{} identifies which object-dependent assets require changes and proposes modifications for Leila to accept or decline before any audio is replaced (\textbf{DG2, DG4}).
For example, making Sophie more nervous rewrites her dialogue to sound more anxious and updates her sound effects (\eg quick and uneven footsteps, door opening slowly with creaking noise) across the full audio drama.
\medskip

\noindent The \textbf{Script Editor} visualizes the audio drama as an annotated script with segments grouped by scene.
Each scene segment consists of dialogue rows, sound effects, and music cues in script order, where any element can be inserted, moved, edited, removed or re-assigned to a different object (\textbf{DG4}).
Voice acting directions can be added inline to any dialogue line (\eg~\audiotag{angry}, \audiotag{whispered}), and multiple lines can be edited at once via a natural-language instruction with per-line system edit suggestions.
For example, Leila adds a \textit{``medieval folk music''} tag to open the market square scene, inserts a paper-crackling sound effect before Martha's line, and rewrites a line with acting directions: ``\textit{\audiotag{angry} Hey! \audiotag{sad} But I wanted to keep talking to you!}''
The Script Editor and Object Pane are linked, so that hovering over an asset in a vignette scrolls the Script Editor to that element and vice versa.
It is also linked to the audio drama playback so that the currently playing word or element is always highlighted and in focus.
\medskip

\noindent The \textbf{Timeline Editor} visualizes the full composition as clips on four tracks (speech, sound effects, music, ambience) and provides clip-level controls inspired by existing timeline-based editors such as placement, fade-in and fade-out, volume adjustment, and trimming and looping (\textbf{DG3, DG4}).
Any clip can be listened to in isolation or regenerated directly from the timeline, and both Script and Timeline Editor are synchronized so that edits in one are immediately reflected in the other.
Leila adjusts timings and volumes for a final mix, then exports the composition as an \textit{.aaf} file to finish mixing and mastering in her preferred DAW.

\begin{algorithm}[t]
\caption{Automatic Audio Drama Production}
\begin{algorithmic}[1]
\State \textbf{Input:} book text $T$
\State \textbf{Extract objects:} $\mathcal{C}, \mathcal{S} \leftarrow \text{LLM}(T)$ \hfill {\scriptsize\color{gray} characters, scenes}
\State \textbf{Instantiate attributes:} $\mathcal{C}, \mathcal{S} \leftarrow \text{LLM}(T, \mathcal{C}, \mathcal{S})$ \hfill {\scriptsize\color{gray} voice, ambience design}
\State \textbf{Generate scripts:} $\mathcal{P} \leftarrow \{\text{LLM}(T, \mathcal{C}, \mathcal{S})\}^{25}$
\State \textbf{Score scripts:} $r(p) \leftarrow \frac{1}{3}\sum \text{LLM}_{\text{judge}}(p)$ \hfill {\scriptsize\color{gray} 3 judges, 11-item rubric}
\State \textbf{Select script:} $p^* \leftarrow \arg\max_{p \in \mathcal{P}}\ r(p)$
\State \textbf{Link assets:} $\mathcal{L} \leftarrow \text{Link}(p^*, \mathcal{C}, \mathcal{S})$ \hfill {\scriptsize\color{gray} speech, SFX, music, ambience $\mapsto$ objects}
\State \textbf{Generate assets:} $\mathcal{A} \leftarrow \text{AudioGen}(\mathcal{L}, \mathcal{C}, \mathcal{S})$ \hfill {\scriptsize\color{gray} using object attributes}
\State \textbf{Arrange \& mix:} $\mathcal{D} \leftarrow \text{Mix}(\mathcal{A})$
\State \textbf{Output:} multi-track audio drama $\mathcal{D}$
\end{algorithmic}
\end{algorithm}

\begin{figure}[t]
    \centering
    \includegraphics[width=\linewidth]{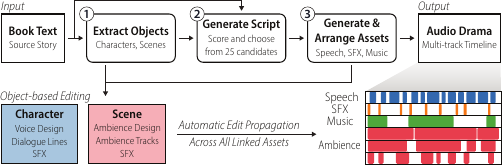}
    \caption{\system{} takes a book as input, then (1) extracts all objects (characters, scenes) and initializes them with attributes (\eg voice and ambience design), (2) generates 25 script candidates and scores them to choose the best based on a rubric derived from audio drama script best practices, and (3) generates and arranges all assets on timeline tracks to output a multi-track audio drama.}
    \label{fig:system-design}
\end{figure}

\subsection{Automatic Audio Drama Production}
\system{} automatically converts raw book text or audio drama scripts (\ie input) into a structured, multi-track audio drama (\ie output) via three steps: (1) extracting objects (\ie characters, scenes), (2) generating and selecting an adapted audio drama script, and (3) generating all audio assets and placing them onto a multi-track timeline (Figure~\ref{fig:system-design}).
\medskip

\noindent \textbf{Extracting \textcolor{speechcolor}{Character} and \textcolor{scenecolor}{Scene} objects.}
Given a raw book text, an LLM analyzes the source text to extract characters and scenes with inferred attributes based on \system{}'s object model (Figure~\ref{fig:system-design}.1).
For each character, \system{} infers a voice design attribute describing the character's vocal traits (\eg \textit{``a middle-aged woman with a measured cadence and a slight regional accent''}), which determines the character's voice across the audio drama (\textbf{DG2}).
For each scene, \system{} infers an ambience design attribute describing the scene's acoustic environment (\eg \textit{``a busy market square with a street band playing''}), which determines the scene's soundscape whenever it is active in the audio drama (\textbf{DG2}).
In addition, \system{} keeps track of dialogue lines in the source text and assigns them to their respective characters derived from their surrounding context.
\medskip

\noindent \textbf{Creating an Audio Drama Script.}
To adapt the book into a script that follows audio drama conventions, we adopt a best-of-$N$ generation strategy in which $N$ candidate scripts are generated in parallel and the highest-quality script is selected using an LLM-based judge~\cite{hashemi_llm-rubric_2024, gu_survey_2024} (Figure~\ref{fig:system-design}.2).
We set $N{=}25$ based on early empirical experiments.
We use an LLM to generate each candidate using a prompt based on the best practices we found in our formative work (Table~\ref{tab:formative_results}).
For example, the prompt instructs the LLM to remove expository narration and convert it into dialogue, externalize internal thoughts as whispered lines, replace concrete character actions with sound effects, insert voice acting directions based on emotional delivery described in the source text, and remove redundancy between characters' speech and their sounds (\textbf{DG1}). For example, the source text \textit{``Martha slammed the cup down on the counter. `I told you not to come here,' she said angrily.''} is converted into:

\begin{quote}
\textcolor{sfxcolor}{\texttt{<sfx prompt="ceramic cup slammed hard on wooden counter" duration="1">}}\\
\textcolor{speechcolor}{\texttt{[VOICE: MARTHA]}} \texttt{"\audiotag{angry} I told you not to come here."}
\end{quote}

\noindent The narration is removed because the sound effects already convey the action, the dialogue is kept but shortened, and the emotional delivery described in the book text (\textit{``she said angrily''}) becomes a voice acting direction (\audiotag{angry}).
Because not every LLM generation will reliably produce a script that follows all audio drama best practices, we score all 25 candidates using an LLM-as-a-judge approach~\cite{hashemi_llm-rubric_2024, gu_survey_2024}.
We score each script candidate three times using an 11-item rubric derived from our formative work best practices (Table~\ref{tab:rubric}), then average the scores for reliability.
We select the script candidate with the highest mean score as the final script.
\medskip

\noindent \textbf{Generating Audio Assets and Assembling the Timeline.}
We use the selected script and the extracted character and scene objects to generate all audio assets (Figure~\ref{fig:system-design}.3).
For each character, we synthesize all dialogue lines using a text-to-speech model and the character's voice design attribute.
To ensure voice consistency across the full audio drama, we concatenate all of a character's dialogue lines into batches and synthesize them together in a single model call, then split them again using word-level alignment timestamps (\textbf{DG2}).
We generate 4 variations for each sound effect using a text-to-SFX model and select the first variation by default to be used in the audio drama.
Because speech represents the main track in audio dramas, we anchor all music and ambience to word-level timestamps of the scripted dialogue to ensure that music and ambience always align with the dialogue they accompany, even when speech durations vary across multiple generations (\textbf{DG3}).
We use an LLM to decompose each scene's ambience design attribute into individual ambience sound layers (\eg \textit{``city hum''}, \textit{``church bells''}, \textit{``distant thunder''}) using a minimized version of Schafer's taxonomy~\cite{schafer_soundscape_1994} as used in SoundStager~\cite{yoo_soundstager_2026} but adapted to the audio drama medium.
For example, we remove signal sounds (\ie foreground sounds, designed to attract attention) as they are more suitable for videos where viewers can see the corresponding visuals, but in audio dramas, they create distractions within scenes and concrete actions already exist as sound effects.
We generate each ambience layer using the same text-to-SFX model to allow editing or removing individual layers within scenes (\textbf{DG2, DG4}).

All generated assets are assembled into a four-track timeline (speech, sound effects, music, ambience), with scene buffers (\ie first and last 5 seconds of a scene only play ambience) and crossfades to create smooth transitions between scenes (\textbf{DG1}).
We ensure that dialogue remains intelligible as the primary carrier of narrative information (\textbf{DG3}) by placing speech at the reference level, with sound effects, music, and ambience at progressively lower volumes, following existing audio drama mixing practices (Table~\ref{tab:formative_results}).
We apply sidechaining (\ie reducing volume when speech is present) to scene ambience and music whenever characters are speaking to prevent other sounds from masking speech (\textbf{DG3}).
We normalize all audio and apply a peak limiter based on audiobook standards~\cite{acxstandard} (\textbf{DG3}).

\subsection{Implementation}
We implemented our pipeline and backend in Python using Flask.
We use Gemini 3 Pro~\cite{googlegemini} for all LLM calls.
All speech is synthesized via ElevenLabs' \texttt{eleven\_v3} model~\cite{elevenlabsv3} using their word-level alignment for per-line slicing.
All sound effects and music are generated through proprietary text-to-SFX and text-to-music models.
All audio is normalized to $-16$\,LUFS and mixed with per-track default volumes for speech (0\,dB), sound effect ($-6$\,dB), music ($-14$\,dB), and ambience ($-20$\,dB) based on best practices (Table~\ref{tab:formative_results}).
We implemented our frontend using React and used the Web Audio API for real-time audio playback.

\section{User Study}
We conducted a user study with 8 professional audio drama producers to compare audio drama creation using object-based audio editing in \system{} to their existing tools and investigate two main questions:
\begin{itemize}
    \item How does \textit{object-based audio editing} support audio drama creation compared to professionals' existing practice?
    \item What are professionals' perspectives on ~\system's automatically generated audio dramas?
\end{itemize}

\subsection{Method}
We used a within-subjects design in which each participant created audio dramas from two different book excerpts, one per condition: \system{} and their existing tools.
To isolate the effect of the production workflow and editing interface, we ensured participants using their existing tools had access to the same speech, and similar sound effects, and music generation models used by \system{}.
Book excerpts averaged 1402 words ($\sigma = 382$), with 5.7 characters ($\sigma = 1.8$) and 4.7 scenes ($\sigma = 2.6$).
Participants always completed the existing-tools condition first, followed by \system{}.

\subsubsection{Participants}
We recruited 8 participants ($7$ male, $1$ female, ages $20$ to $56$) from Upwork~\cite{upwork} (Table~\ref{tab:participants}).
All participants had prior audio editing experience ($\mu = 6.1$ years, $\sigma = 5.8$ years), created or edited audio dramas or immersive audio storytelling content before, and frequently use generative audio tools in their existing workflows (\eg ElevenLabs~\cite{elevenlabsv3}, Suno~\cite{suno}, Stable Audio~\cite{evans_stable_2024}).
We compensated participants at their hourly Upwork rate (\$25-\$40).

\subsubsection{Procedure}
We ran a 2-stage study.
In Stage 1, participants created an audio drama using their existing tools.
In Stage 2, participants used \system{} in a moderated session.
\smallskip

\noindent \textbf{Stage 1.}
Participants completed a demographic questionnaire, reviewed audio drama reference examples\footnote{\url{https://www.audible.com/blog/article-best-classic-audio-dramatizations}}, and were given a book excerpt with the goal of creating a compelling, engaging, and immersive audio drama within five business days using their own tools.
For each step of their workflow, participants tracked the time spent and wrote a short report describing their process.
\smallskip

\noindent \textbf{Stage 2.}
In a 100-minute moderated session, participants began with a 20-minute guided tutorial of \system{} using a sample project.
Then, participants had 40 minutes to create an audio drama from a new book excerpt following the same goal as in Stage 1, followed by an unlimited time to refine their audio within 5 business days (similar to the existing tools condition).
We measured total \system{} production time from the start of the creation task to the completion of final mixing and mastering.
Participants completed a questionnaire comparing both \system{} and their existing tools using 5-point Likert scales drawn from the NASA-TLX~\cite{hart_development_1988}, the Creativity Support Index~\cite{cherry_quantifying_2014}, and the System Usability Scale~\cite{brooke_sus_1996}, followed by a 20-minute semi-structured interview.

\subsubsection{Study limitations}
Participants always completed the existing-tools condition before \system{}.
This ordering introduces two opposing effects as participants may have benefited from task-level familiarity when using \system{} second, but were also disadvantaged by having years of experience with their own tools compared to only a 20-minute introduction to \system{}.
We screened participants for prior audio drama experience and assigned different book excerpts between conditions to reduce carry-over effects.
Participants further had unlimited time using their existing tools to complete Stage 1 and for the final mixing and mastering at the end of Stage 2, but there was a 40-minute in-session limit using \system{} during Stage 2, which may underestimate what participants could accomplish with \system{} given more time.
Participants' time estimates were self-reported and should only be used as complementary signals to make conclusions about time savings.
Lastly, while common in HCI research, our sample of 8 professionals may limit the generalizability of our findings.

\begin{figure*}[t]
    \centering
    \includegraphics[width=\linewidth]{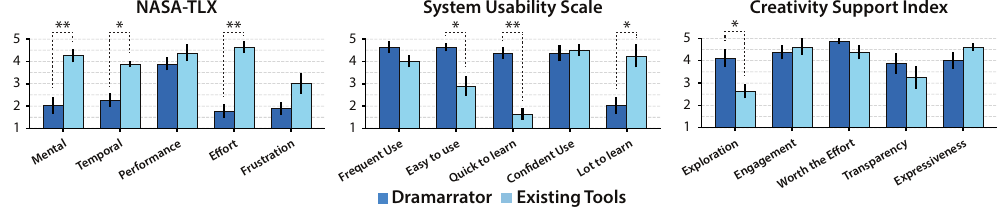}
    \caption{Participant ratings ($N=8$) for questions selected from NASA-TLX, System Usability Scale, and Creativity Support Index comparing \system{} and their existing tools. Error bars represent a 95\% confidence interval. *\alphaval{<}{.05}, **\alphaval{<}{.01}}
    \label{fig:study-quants}
\end{figure*}

\subsection{Results}
All participants preferred the object-based editing paradigm of \system{} over their existing tools because \system{} significantly reduced the task load and effort required to create audio dramas.

\medskip
\noindent \textbf{\system{} reduces task load and is easier to use than existing tools.} Participants reported significantly lower mental demand, temporal demand, and effort when working with \system{} compared to their traditional tools (\alphaval{<}{.05}, Figure~\ref{fig:study-quants} left).
Participants rated ~\system{} as significantly easier to use and faster to learn compared to their existing tools (\alphaval{<}{.05}, Figure~\ref{fig:study-quants} middle).
Participants rated ~\system{} significantly higher on exploration (\alphaval{<}{.05}) while engagement, transparency, expressiveness, and perceived worth of effort remained comparable between conditions (Figure~\ref{fig:study-quants}, right).
We analyzed task load, creativity support, and usability ratings using paired Wilcoxon Signed-Rank tests~\cite{woolson_wilcoxon_2007}.
We report all means, standard deviations, and significance values in Table~\ref{tab:ratings_significances}.

\medskip
\noindent \textbf{Object-based editing amplifies creative control over characters and scenes.}
Seven out of eight participants used object-level editing features during their sessions with an average of $2.6$ ($\sigma=1.4$) object-based edits that propagated into $25.0$ ($\sigma=20.5$) manual edits, a $9.7\times$ amplification in editing effort. 
The amplification was highest for characters with many dialogue lines: P4's single voice edit on Mr.\ Utterson propagated to all 30 of his speech lines, and P5's voice edit on Mr.\ Prosser propagated to all 18 lines.
P6 reimagined the narrator's voice from ``\textit{an old man's voice, sounding like a storyteller}'' to ``\textit{a mysterious, low-toned narrator, whispering secrets of a lost kingdom}'' which P6 described as \textit{``a single creative decision that automatically replaces everything across the audio''}.
P8, the most experienced participant (17 years) used object-based editing most extensively, for example, by iterating on 10 voice designs across 4 characters that saved 56 manual edit operations in total:
\begin{quote}
    ``\textit{changing voices would normally take an hour, and now this was 10 seconds [...] if I do this the old school way I have to manually do that on the timeline which would take me 30 minutes and here I can just change that.}''
\end{quote}

While P3, who did not make edits to objects, expressed skepticism that a single AI voice prompt for a character produces consistent results across all lines, and prefers a line-by-line workflow, other participants' workflows aligned better with \system's object model.
P1 stated the object-based approach ``\textit{definitely should be a product --- the tedious, annoying work is automated}'', P4 found ``\textit{having the ability to break down scenes and have the control --- that's brilliant}'', and P7 highlighted that the object-based approach ``\textit{allows me to make the story more alive and create contrast between the characters}''. P5 contrasts their existing tools to object-based editing:
\begin{quote}
    ``\textit{the objects are the diamond in the tool [...] [changing a character's voice] would take forever and would take weeks especially in a full production with voice artists who then have to re-record [...] that would be almost impossible if you already have a whole book in your editor and you just want to change a character, that would be like redoing the whole project all over again.}''
\end{quote}

Participants used object-based editing to support creative exploration (Figure~\ref{fig:study-quants}, right).
Our interaction logs show that participants spent $60\%$ ($\sigma=16\%$) of their editing time in the script editor, $27\%$ ($\sigma=14\%$) in the object pane reviewing character and scene objects, and only $9\%$ ($\sigma=7\%$) in the timeline view.
The majority of object pane interactions ($\mu=138.4$, $\sigma=72.9$ in total) were browsing character and scene vignettes ($\mu=110.8$, $\sigma=60.8$), followed by previewing audio from the vignettes ($\mu=17.1$, $\sigma=12.7$), then object-level attribute edits ($\mu=10.5$, $\sigma=7.2$).
Five participants (P1, P2, P6, P7, P8) started their sessions by browsing character and scene vignettes in the sidebar to get an overview of the generated audio drama before making any edits.
Participants repeatedly revisited the same objects throughout their sessions, returning to each unique character an average of $17.3$ ($\sigma=11.1$) times and each scene $6.7$ ($\sigma=4.9$) times, and frequently switching between different characters to compare and ensure contrast between voices.
P4 noted: ``\textit{just the overviews of the objects, I think that's really good --- having the ability to break down scenes and have the control --- that's brilliant}''.
Participants also used the sidebar to compare characters against one another: P7 highlighted character objects ``\textit{allow me to make the story more alive and create contrast between the characters}.'' 

\begin{figure}[h]
    \centering
    \includegraphics[width=\linewidth]{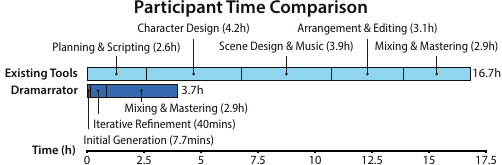}
    \caption{Self-reported time breakdown for existing tools and \system{}. \system{} reduces audio drama creation time by a factor of 4, from 16.7 ($\sigma=7.5$) to 3.7 ($\sigma=2.3$) hours.}
    \label{fig:study-times}
\end{figure}

\medskip
\noindent \textbf{Exploratory analysis of time savings.}
We complement participants' perceived efficiency gains by comparing self-reported times between the existing tools and \system{} condition as supporting signals.
\system{} reduced the time to create audio dramas by a factor of 4, from 16.7 ($\sigma=7.5$) to 3.7 ($\sigma=2.3$) hours (Figure~\ref{fig:study-times}).
With their existing tools, participants spent their time across 5 sequential phases: planning and scriptwriting ($\mu=2.6$\,h), voice creation and dialogue production ($\mu=4.2$\,h), sound effects and music creation ($\mu=3.9$\,h), timeline arrangement ($\mu=3.2$\,h), and mixing and mastering ($\mu=2.9$\,h).
Each phase required different tools (\eg ElevenLabs~\cite{elevenlabsv3} for speech, sound effects libraries~\cite{epidemicsound}, Suno~\cite{suno} for music, and editing tools like Adobe Audition~\cite{adobeaudition}), and changing an earlier decision such as a character's voice required starting over from that phase.
With ~\system{}, the automatic audio drama generation pipeline handled scriptwriting, voice casting, sound design, and composition automatically in 7.7 ($\sigma=1.6$) minutes on average, after which participants refined the result in the editing interface for approximately 40 minutes, followed by 2.9 hours of final mixing and mastering in their existing tools.
All participants described the generated output as a strong starting point: P2 called it ``\textit{almost all the way done}'', P5 summarized it as ``\textit{a sketch that I can then enhance [...] with small adjustments this could be turned into a great audio experience}'', and P7 highlighted that ``\textit{the emotions were right, it had actual characters with personalities}'', and P8 noted ``\textit{it really kept the script well, the writing was not cheesy}''.
All participants highlighted the time savings: P1 estimated ~\system{} ``\textit{saves you like 4--5 hours}'', P2 noted it ``\textit{might save me a day of work}'', P3 estimated ``\textit{5 hours just to get the initial composition}'', P4 described the savings as ``\textit{a lot -- hours, if not days}'', P5 reported it ``\textit{saved me more than 15 hours}'', P7 estimated the process to be ``\textit{70\% faster}'', and P8 noted ``\textit{manually creating this audio drama took me 10 hours or more, now this would just take me an hour}''.

\medskip
\noindent \textbf{\system's workflow implications.}
Object-based editing shifted participants from a sequential workflow that consists of pre-production (\eg scripting, planning characters and scenes), production (\eg recording or generating assets), and post-production (\eg assembling and editing assets on a timeline) to an iterative workflow that enabled more experimentation. For example, P8 noted they could make ``\textit{creative changes on the fly}'', a workflow that P6 described as:
\begin{quote}
    ``\textit{[\system] allows me to listen, change something, listen, change something --- otherwise I have to manually create everything first and then it's hard to change something}''
\end{quote}

Participants also used character and scene objects to shape their narrative depth within the audio drama.
For example, P7 gave one character a rougher, lower voice to contrast with another, brighter-voiced character, switching back and forth between editing the voice design and listening to a dialogue sequence between the two to check that they were now clearly distinguishable.
P7 described this process as a way to ``\textit{add depth to each character [...] and make the story more alive and create contrast}.'' 
P8 described iteratively designing characters to add distinctive sonic ``\textit{color}'' to the system-generated result: ``\textit{I can really be designing these scenes and characters and then always reuse them.}''

\medskip
\noindent \textbf{Future use cases for \system{}.}
All participants expressed that they want to use ~\system{} for future projects.
P3 and P4 preferred ~\system{} as a plugin integrated in their DAW rather than a standalone tool, while P8 suggested it could also work as standalone software to cater to non-audio-engineers such as publishers and authors.
Participants mentioned use cases for \system{} beyond converting books to audio dramas such as game design (P1, P2, P4, P6, P7, P8), immersive audio stories (P1, P3, P4), scripted podcasts (P4, P8), YouTube videos with generated audio (P1, P7), animations and trailers (P5, P8), and radio and commercials (P3).
P5 envisioned a use case for film where \system{} may be used to create a higher-fidelity storyboard to better visualize film sequences (``\textit{like a table read [...] basically an audioboard}'').

\medskip
\noindent \textbf{Participants' concerns regarding AI audio generation and perspectives on automation.}
The most common concern raised across participants was AI audio generation quality.
Participants highlighted acoustic inconsistency in AI-generated audio assets (speech, sound effects, music) that sometimes sound like they were recorded in different environments, and thus require post-processing such as EQ, reverb matching, and de-essing to unify the acoustic space.
Participants were also sometimes frustrated by unreliable sound effect generation that required multiple regenerations to receive a satisfactory result.
To address these issues, participants used object-based editing to fix voices at the character-level or remove sounds with artifacts from scene soundscapes via the scene vignettes, rather than having to manually locate and replace each affected clip individually.
Participants requested improvements such as connecting audio engineering aspects (\eg reverb, spatial mixing) to object-based edits so that scene-level acoustic environments propagate automatically (\eg P8 describes how a cave scene should apply the same reverb to all characters within that scene), chat-based prompt interactions to refine the overall script (P5), more distinction between generation variations (P2, P8), and finer timeline controls such as gap and pacing adjustments (P1, P5).
However, all participants noted that being able to export the multi-track project file allows them to repair these issues in their existing tools.

Despite these quality concerns, participants viewed ~\system{} as automating tedious production tasks while keeping creative decision-making in their hands.
P5 noted that ~\system{} gives them ``\textit{the opportunity to be the director}'' rather than a technician, and P8 described the tool as filling the role of a production assistant --- a role P8 typically hires for --- freeing them to focus on creative work and final mixing.
P1 similarly valued that ``\textit{the tedious, annoying work is automated}'' while retaining full control over creative choices.
A secondary concern expressed by some participants (P1, P8) was long-term job displacement by AI automation, though this was framed in the context of broader industry trends rather than ~\system{} specifically.

P8 also raised a concern about artistic homogenization and thus AI tools removing individual creative distinctiveness: ``\textit{if everybody uses that tool there might be a tendency that everyone's [audio] sounds the same, so I might want to add my own texture.}''
P3 raised a concern about copyright and questioned the legal ownership of AI-generated audio when using generated assets in their professional work.

\begin{figure*}[t]
    \centering
    \includegraphics[width=\linewidth]{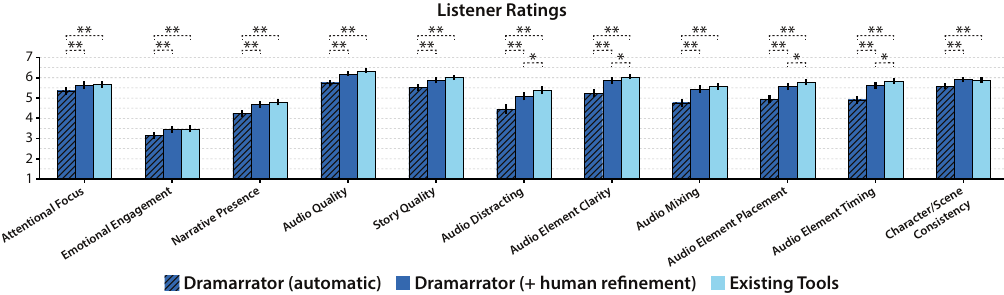}
    \caption{Listener ratings ($N=300$) comparing audio dramas across three conditions: automatically generated by \system{}, refined by professionals using \system{}, and produced by professionals using their existing tools. Error bars represent a 95\% confidence interval. *\alphaval{<}{.05}, **\alphaval{<}{.01}}
    \label{fig:result-quants}
\end{figure*}

\subsection{Listener Evaluation}
The user study above helped us understand creators' perceptions of \system{} and its output quality. To complement these findings from the audience side, we further evaluate the output quality through a listening study. Specifically, we collected ratings from 300 listeners on Prolific~\cite{noauthor_prolific_nodate} comparing the raw output from our pipeline (\textit{\system{}-Auto}), the versions that creators refined based on the pipeline output (\textit{\system{}-Refine}), and the golden examples that creators created using their existing tools during Stage 1 of the user study (\textit{Existing-Tools}).
We randomly chose 6 stories from our user study.
Each listener was assigned three stories and rated one randomly chosen condition per story.
To assess different dimensions of listener experience, we derived attentional focus, emotional engagement, and narrative presence from the User Engagement Scale~\cite{busselle_measuring_2009}, and asked listeners to rate overall audio and story quality, five audio element arrangement and mixing questions, and character and scene consistency on 7-point Likert scales.
We analyzed ratings using Friedman tests with pairwise Wilcoxon tests and Bonferroni correction (Figure~\ref{fig:result-quants}). Below, we summarize the key findings.

\smallskip
\noindent\textbf{Human refinement helps bridge the gap to professional quality.}
Listeners rated \textit{Existing-Tools} higher than \textit{\system{}-Auto} on all 11 measures (\alphaval{<}{.01}).
However, \textit{\system{}-Refine} significantly improved \textit{\system{}-Auto} on all 11 measures (\alphaval{<}{.01}), achieving ratings comparable to \textit{Existing-Tools} on 7 measures, including user engagement, overall audio and story quality, and character and scene consistency.
Notably, character and scene consistency was rated on par with professionally produced audio dramas.
Object-based audio editing in \system{} likely has supported maintaining consistency, as attribute changes to character or scene objects automatically propagate to all dependent assets, compared to manual clip-by-clip timeline editing in existing tools.

\smallskip
\noindent\textbf{A gap remains in audio element distraction, clarity, placement, and timing.} Despite the significant improvement after human refinement, listeners still rated \textit{Existing-Tools} higher than \textit{\system{}-Refine} on distraction, clarity, placement, and timing of audio elements (\alphaval{<}{.05}). We identified two potential sources for this gap: (1) AI generation artifacts (\eg unnatural sound effects) that may go unnoticed during refinement but become apparent to listeners, and (2) constraints of \system{} (\eg uniform pauses between speakers, no sound effects overlapping with speech).

\section{Exploratory Study with Storytellers Beyond Audio Drama}
While \system{} was designed with audio drama creators in mind, we were also interested in whether the tool could lower the barrier to entry enough that storytellers without audio production expertise might consider creating immersive audio stories, and whether the principles behind \system{} may generalize to other domains of storytelling. To this end, we invited three participants (Table~\ref{tab:participants-exploratory}) to a 90-minute session in which they explored \system{} while discussing its usefulness and the potential of object-based editing within their creative domains: an escape room designer (N1), a tabletop RPG enthusiast (N2), and a computer game designer (N3).

All three found \system{} easy to use and expressed interest in using \system{} to add immersive audio stories to their own work. While professionals valued \system{} primarily for the time it saved, novices valued that it made audio drama creation accessible to them at all.
N3 noted ``\textit{I definitely would not be able to create audio dramas without experience but now I could}''. All  three participants connected object-based editing to pain points in their own domains.
N1 described how a client requesting a character change in an off-the-shelf escape room they sell, currently requires ``\textit{basically creating a fully new project doing everything manually}''.
N2 noted that tabletop RPG game masters often need to adapt scenarios on the fly, and suggested that object-based editing could allow real-time swapping of character voices, scene ambience, or dialogue tone. 
N3 observed that high-level creative changes in game production grow increasingly costly the later they occur, since every dependent asset must be updated manually.
This suggests that object-based editing addresses a broader challenge in narrative content creation that extends beyond audio drama production.

\section{Discussion and Future Work}
\label{sec:discussion}
Our evaluations show that \system{} addressed our design goals.
In our user study, \system{} lowered professionals' task load and \textit{object-based audio editing} enabled iterative shaping of characters and scenes (\textbf{DG2}, \textbf{DG4}).
A listener evaluation revealed that automatically generated audio dramas refined via object-based audio editing in \system{} were comparable to those created by professionals using their existing tools on engagement, audio and story quality, and character and scene consistency measures (\textbf{DG1, DG2, DG3}).
We reflect on our findings and discuss future opportunities to extend and generalize \textit{object-based audio editing} in \system{}.

\subsection{Extending the Object Model}
\system{}'s object attributes currently cover narrative properties (\eg voice design, ambience design), but professionals in our user study requested extensions that would deepen the range of edits a single object change can capture.
Participants requested acoustic properties as object attributes.
For example, P8 described how a cave scene should automatically apply the same reverb to all characters speaking within it, rather than requiring manual post-processing for each asset.
Just as a character's voice design attribute currently defines their vocal identity and propagates to all their speech assets, scene objects could carry an acoustic environment attribute (\eg room size, reverb) that is applied to all assets linked to that scene.

We can apply \system{} to arbitrarily long audio dramas up to the size of the LLM’s context window. However, our object model enforces consistency of characters and scenes, which may change over the course of a story (\eg a character may age).
Borrowing object inheritance from object-oriented programming~\cite{rentsch_object_1982, stefik_object-oriented_1985} would allow the creation of multiple instances of the same object (\eg child Alice, adult  Alice) that share base attributes but differ in attributes based on the story plot (\eg adult Alice's voice is heavier).
Future iterations could also extend the usage timeline in \system{}'s object vignettes with a visualization that shows when attributes are active, and let creators drag attribute ranges to specific story segments, similar to animation tools~\cite{aftereffects, blender, toonboomharmony} where creators define when a property applies via keyframes.
For relationship-based conditions (\eg a character's tone shifting depending on who they are with), an object could hold conditional attributes that activate based on other objects that co-occur in the same scene.

\system{} currently relies on an LLM's latent knowledge to translate high-level object edits across asset modalities (\eg how a character's heightened anxiety should  modulate their speech, accompanying Foley, and background music simultaneously).
While flexible, this approach lacks transparent formal logic 
that creators can inspect or override.
To make this mapping more interpretable, future systems could implement a cross-modal mapping that is grounded in a shared dimension.
For example, prior work showed that valence and arousal~\cite{russell_circumplex_1980} can be used as an emotional dimension that maps onto individual modalities, such as text~\cite{juslin_vocal_2005}, speech~\cite{preotiuc-pietro_modelling_2016}, 
music~\cite{rubin_generating_2014}, and sounds~\cite{fan_emo-soundscapes_2017}.
This suggests there may be a shared emotion space that could inform a cross-modal mapping to handle dependencies.
For example, after a creator makes edits to a scene and therefore shifts its emotional valence, a system could identify all assets which are now out of alignment with the shift in valence to inform what assets need to be updated, rather than relying on an LLM.
However, regenerating assets to achieve a desired emotion 
shift remains an open challenge.

\subsection{Generalization of Object-Based Editing}
Our informal study with 3 novices from adjacent creative domains revealed that the object-based editing paradigm may generalize beyond audio drama production.
All three highlighted that their existing tools force them to edit each dependent asset individually, even though they already conceptualize their creative content through characters, scenes, and environments.
These observations suggest that wherever a creative domain is organized around narrative constructs with many dependent assets, reifying those constructs as editable objects with propagating attributes~\cite{beaudouin-lafon_instrumental_2000, xia_object-oriented_2016} could provide a similar effort amplification.
For example, in video production, character and scene objects could propagate changes across dependent visual sub-channels such as lighting, camera framing, or character expressions, where the main challenge compared to audio is that regenerating visual assets risks introducing inconsistencies with surrounding frames. 

\subsection{Implications for Collaborative Work}
While \system{} was designed for individual creators to reflect an emerging practice of individuals expanding their ability to create with AI~\cite{daugherty_human_2024, louie_novice-ai_2020, luo_designing_2025}, we see the potential for object-based audio editing to also expand collaborative workflows.
Even without AI asset generation, the object model could serve 
as a coordination layer among human collaborators. For example, character objects could  export per-character casting briefs for voice directors, or scene objects could track which assets are pending or delivered.
In the future, we can also enable real-time collaboration in \system{} by locking assets linked to objects that are currently modified to prevent edit conflicts.
Lastly, instead of replacing voice actors, who offer qualities AI cannot simply replicate~\cite{jiang_forging_2025}, \system{}'s AI-generated audio drama outputs can serve as high-fidelity prototypes that support creators to concretize story, character, and scene decisions before human talent is engaged --- similar to preliminary ``table-read'' sessions before final studio recordings~\cite{proferes_film_2005, leach_theatre_2013}.

\subsection{Creative Agency and Ethical Implications}
Professionals described \system{} as freeing them to act as creative directors rather than manual assemblers, because the system handles low-level production so they can focus on shaping characters and scenes.
Producers can iterate through many character designs or scene atmospheres and select the best, rather than committing early to save revision costs.
For example, P8 iterated over 10 voice designs across 4 characters in a single session, a workflow that previously required hours of manual timeline work.

Handing low-level production to the system also increases the abstraction distance from the concrete assets, which introduces new tensions.
Some participants found higher-level control liberating (P4, P6, P8), while others found the added flexibility disorienting or felt less ownership over the result (P4, P8).
This reflects a known tension in AI-assisted creativity tools between automation that amplifies creative capacity and automation that shifts creative authorship toward the system~\cite{li_beyond_2023}.
\system{} shows this tension at the level of object attributes. A character's voice design or a scene's ambience is a natural-language prompt that determines how the object appears across the project, yet creators currently cannot inspect or override the normative design guidelines that \system{} uses to infer such prompts from the book text (Table~\ref{tab:formative_results}).
In the future, we should let creators edit, extend, or replace these guidelines for more control.
Object-based editing also shifts the creator's effort from making each edit to reviewing the edits the system proposes, and creators should be able to decide how much review each propagation deserves.
For edits where ownership matters, we can introduce selective good friction~\cite{chen_exploring_2024, cox_design_2016} to increase creators' sense of ownership, for example by requiring creators to write the regeneration prompt for each affected asset themselves so that the review effort stays meaningful.
For routine edits, we can instead reduce the review burden, for example by learning from previously accepted edits to auto-approve similar ones.

\system{} uses generative AI tools, such as LLMs~\cite{googlegemini} and audio generation models~\cite{elevenlabsv3, adobefirefly}, which carries creative potential but also ethical concerns.
Generative AI tools can augment the creative process by turning creators' intents into assets almost instantly, and thus significantly speeding up the time between instantiation, assessment, and refinement of ideas~\cite{tseng_keyframer_2024, riche_ai-instruments_2025}.
Furthermore, generative AI tools lower barriers to content creation and often make creating content accessible in the first place~\cite{wessel_generative_2025, constantinides2018introduction, huy_generative_2024, pinski_ai_2024}.
For example, creating audio dramas traditionally required specialized expertise and significant resources, but our studies showed \system{} lowers barriers that once limited who could take part in creating audio drama content.
Lowering barriers and automation also raises concerns about long-term job displacement from AI automation, which were also voiced by some participants in our user studies (P1, P8).
For example, P8 described using AI in place of a production assistant they used to hire for parts of their work.
P3 and P8 also worried about who owns content created this way.
P3 questioned the copyright and legal ownership of AI-generated audio used in professional work, because P3 felt the audio was \textit{not their own work}.
P8 worried that if many creators use the same AI tools, audio content could start to sound alike, so creators may need to add their own texture to stay distinctive.

\section{Conclusion}
We presented \system{}, an audio drama authoring system that introduces object-based audio editing to support creators in adapting books into audio dramas. \system{} extracts narrative constructs (\eg characters, scenes) from book text or original scripts as objects with attributes (\eg a character's voice, a scene's ambience), automatically generates a multi-track audio drama, and links each asset to its object so that high-level edits propagate automatically across all dependent assets.
A user study with 8 professionals showed that \system{} significantly lowered task load compared to existing tools.
Object-based audio editing amplified editing effort by $9.7\times$ on average, with a single object change replacing up to 30 manual edits while keeping characters and scenes consistent across the entire audio.
A listener study (N=300) revealed that automatically generated audio dramas refined via object-based audio editing in \system{} were comparable to those created by professionals using their existing tools on engagement, audio and story quality, and character and scene consistency measures.
Finally, an exploratory study with 3 storytellers without audio expertise showed that \system{} lowers the barrier to audio drama creation and object-based editing may generalize to other domains organized around narrative constructs with cascading asset dependencies.
We hope our work inspires future research into object-based editing abstractions for narrative content creation tools.


\bibliographystyle{ACM-Reference-Format}
\bibliography{references}

\appendix

\clearpage\onecolumn  
\section{Appendix}

\footnotesize
\begin{longtable}{@{}
  >{\raggedright\arraybackslash}p{1.8cm}
  >{\raggedright\arraybackslash}p{3.1cm}
  >{\raggedright\arraybackslash}p{6.4cm}
  >{\raggedright\arraybackslash}p{3.5cm}
  >{\raggedright\arraybackslash}p{2cm}
  @{}}

\caption{Guidelines for creating audio dramas, derived from qualitative content analysis of professional audio drama literature. Guidelines are organized by the four design goals (DG1--DG4) they inform.}
\label{tab:formative_results} \\

\toprule
\textbf{Theme} & \textbf{Goal} & \textbf{How} & \textbf{Why} & \textbf{Sources} \\
\midrule
\endfirsthead

\multicolumn{5}{l}{\tablename~\thetable{} -- \textit{continued from previous page}} \\[3pt]
\toprule
\textbf{Theme} & \textbf{Goal} & \textbf{How} & \textbf{Why} & \textbf{Sources} \\
\midrule
\endhead

\midrule
\multicolumn{5}{r}{\textit{Continued on next page\ldots}} \\
\endfoot

\bottomrule
\endlastfoot

\multicolumn{5}{l}{\textbf{DG1: Produce an effective audio drama script}} \\
\midrule

\textbf{Script Structure \& Pacing}
& Keep scenes short and momentum-driven
& Target individual scenes of under 1--2 minutes of audio ($\approx$\,1--2 script pages) and complete episodes of 25--30 minutes (22--27 pages). After drafting, ask of every scene: \emph{``Does this scene advance the central conflict?''} Merge adjacent scenes if combining them loses nothing. Cut any scene that only recaps prior events without advancing conflict.
& Audio listeners have no visual anchor; a prolonged scene without action or conflict causes attention to drift. Short scenes maintain urgency and momentum.
& \cite{toscan2023writingaudiodrama, wayland2020bombs, yt-bbcscriptwriting, yt-bbcwritingforradio} \\

\addlinespace

& Deliver exposition just-in-time; never front-load it
& Scatter all background information (character history, relationships, world-building) as small fragments within natural dialogue exchanges across multiple scenes. Never open an episode with an expository monologue or narrator backstory dump. Apply the rule: \emph{there are no introductions in dramatic writing}---begin every scene mid-action and let exposition emerge organically from what characters say to each other.
& Front-loading background information overwhelms listeners and kills engagement before the story can grip them; listeners disengage before the drama has a chance to begin.
& \cite{toscan2023writingaudiodrama, wayland2020bombs, yt-bbcscriptwriting, yt-bbcwritingforradio} \\

\addlinespace

& Open each episode with a compelling hook; close to invite re-engagement
& Begin every episode \emph{in medias res}---at a moment of conflict, tension, or mystery---within the first 30 seconds. Do not open with scene-setting narration or pleasantries. End each episode with an unresolved question, a cliffhanger, or a high emotional moment that leaves the listener in suspense. Never use narration to recap the previous episode.
& Audio is a passive medium; listeners can disengage effortlessly. The opening hook secures their commitment; the closing hook drives return engagement across episodes.
& \cite{de_fossard_writing_2005, yt-bbcscriptwriting, yt-bbcwritingforradio} \\

\addlinespace

\textbf{Dialogue Craft}
& Write tight, natural, conversational lines
& Enforce the \emph{right-margin rule}: each spoken line should be no longer than $\approx$\,17 words. Avoid multiple consecutive sentences in one character's turn. Read every line aloud---if it sounds unnatural when spoken, rewrite it. Cut filler phrases, pleasantries, and any line that neither advances the plot nor reveals character. Avoid literary constructions that work on the page but confuse listeners when heard.
& Dense or literary dialogue crowds out the SFX and ambience that make audio drama immersive; it sounds stilted when performed and strains listener comprehension.
& \cite{toscan2023writingaudiodrama, wayland2020bombs, yt-bbcscriptwriting, yt-bbcwritingforradio} \\

\addlinespace

& Limit each scene to two characters (\emph{two-handers})
& Design every scene as a dialogue between exactly two characters. When a third character must briefly appear, give them a single functional line and remove them immediately. Audit the entire script and cut any character who: (a) appears for less than one full script page in total, (b) has no direct connection to the central conflict, (c) exists solely as a technical device (\eg a server who takes an order and is never heard from again).
& More than two simultaneous voices creates confusion about character identity in a purely auditory medium, undermining story comprehension.
& \cite{toscan2023writingaudiodrama, wayland2020bombs, crook_radio_2002, yt-bbcscriptwriting, yt-bbcwritingforradio} \\

\addlinespace

& Write for the ear, not the eye
& Favor short, concrete, everyday words over long or abstract ones. Test each sentence by reading aloud at performance pace and mark any phrase that feels awkward or trips the tongue---rewrite it. Prefer active voice and simple sentence structures. Avoid tongue-twisters, complex nested clauses, and multi-part lists within a single sentence. If a sentence requires the listener to hold too many concepts before resolving, break it into two.
& Listeners cannot pause and re-read; every sentence must land on the first pass. Phonetically difficult or structurally complex lines cause comprehension failure that cannot be corrected in a live listening experience.
& \cite{crook_radio_2002, yt-boothjunkieaudiodramasounddesign} \\

\addlinespace

\textbf{Narration \& Prose Adaptation}
& Minimize narration; use it only when no other option exists
& Before accepting any narrator line as final, ask: \emph{``Can this be conveyed through dialogue or sound?''} Retain narration only for four cases where audio has no alternative: (1) character physical appearance, (2) colors, (3) a specific named location that cannot be established through ambience alone, (4) an action that would be genuinely ambiguous without verbal clarification. Cut or convert every other narrator line. No single narrator passage should exceed $\approx$\,30 seconds of spoken audio.
& Narration is described as \emph{``diluted poison''} for audio drama: small amounts are tolerable, but excess stalls narrative momentum and regresses the production toward the audiobook format, destroying immersive potential.
& \cite{wayland2020bombs, toscan2023writingaudiodrama, bernaerts_audionarratology_2021} \\

\addlinespace

& When narration is used, make it intimate and conversational
& Cast the central character as the internal narrator (first-person \emph{homodiegetic}). Write narrator lines as if the character is confiding privately to a trusted listener: use contractions, casual vocabulary, and natural pauses. Avoid formal or literary sentence structures in narrator lines. Keep narrator monologues to 15--30 seconds before returning to scene action. Record the narrator voice dry (no reverb, close-mic) to reinforce intimacy with the listener.
& A formal, literary narrator voice breaks the immersive illusion of a performed scene and reminds the listener they are hearing a narration; an informal narrator creates an intimate bond that sustains engagement.
& \cite{toscan2023writingaudiodrama, bernaerts_audionarratology_2021, yt-cactusaudiodrama} \\

\addlinespace

& Externalize internal monologue and prose description into dialogue or sound
& When adapting a prose source, identify every passage of: (a) internal thought (\eg \emph{``She wondered if he knew''}), (b) descriptive narration (\eg \emph{``The room smelled of old books''}), (c) reported speech (\eg \emph{``He told her he was leaving''}). For each passage, apply one of three conversion strategies: (1) \emph{Dialogue conversion}---rewrite internal thoughts as direct speech between characters or brief first-person narrator asides; (2) \emph{Sound conversion}---replace environmental description with an ambience or SFX annotation (\eg \texttt{[AMBIENCE: dusty, close acoustic space; faint creaking of shelves]}); (3) \emph{Delivery conversion}---encode subtext in the actor's delivery direction rather than in narration.
& Audio drama is a performative, not descriptive, medium. Internal thought and prose description are invisible to listeners and must be externalized into the audible world to exist. This is the central challenge of adapting prose fiction for audio.
& \cite{rattigan_theatre_2002, bernaerts_audionarratology_2021, toscan2023writingaudiodrama} \\

\addlinespace

& Avoid cross-modal redundancy between narration and sound
& After assembling a scene with all SFX and ambience, re-read every narrator and character line while listening to the mix. Flag any line describing an action the listener can already hear (\eg remove \emph{``just a minute, while I open this crate''} if the crate-opening SFX is audible in the mix). Cut or substantially rephrase all flagged lines. Deliberate redundancy---for comic effect or critical emphasis---is only permissible as a conscious, purposeful choice.
& When sound already tells the listener what is happening, narrating the same information creates boredom, wastes scene time, and signals a lack of trust in the listener's intelligence, damaging narrative engagement.
& \cite{wayland2020bombs, toscan2023writingaudiodrama, bernaerts_audionarratology_2021, yt-cactusaudiodrama} \\

\midrule

\multicolumn{5}{l}{\textbf{DG2: Give characters and scenes distinct, consistent acoustic identities}} \\
\midrule

\textbf{Character Design \& Voice}
& Give each character an immediately identifiable voice
& Before scripting, write a one-page voice profile for each character defining: (1) pitch range relative to other characters (high/low), (2) speaking pace (fast, measured, or halting), (3) accent or dialect, (4) vocabulary level (formal, colloquial, or technical), (5) habitual phrases or verbal tics (\eg always beginning sentences with ``Look,\!'' or trailing off mid-sentence). Ensure no two main characters share more than one of these features. Actors must internalize and maintain their profiles throughout the entire production.
& Voice is the sole identifier of character in audio drama; indistinct voices cause listeners to confuse characters and lose track of the story.
& \cite{wayland2020bombs, crook_radio_2002, de_fossard_writing_2005, rattigan_theatre_2002, yt-bbcscriptwriting, yt-bbcwritingforradio, yt-ryanaudiodrama} \\

\addlinespace

& Ensure listeners always know who is speaking
& Use character names frequently and naturally within dialogue---at least once every 3--4 exchanges (\eg \emph{``I hear you, Sarah, but\ldots''} rather than \emph{``I hear you, but\ldots''}). Dedicate a new character's entire first scene to establishing their voice; never introduce a character in a single line. Use recurring story devices to re-anchor identity: a catchphrase, a distinctive sound cue on entry, or a structural habit (\eg one character always asks questions, another always gives commands).
& In a purely auditory medium, name repetition and consistent character habits are the primary anchors for listener comprehension; without them, listeners lose track of who is driving the scene.
& \cite{wayland2020bombs, toscan2023writingaudiodrama, yt-writingcomics} \\

\addlinespace

& Use voice texture to convey personality and emotion beyond the words
& Direct actors on the \emph{physical} qualities of their voice: breath patterns (controlled and even\,=\,authority; ragged and shallow\,=\,fear or excitement), resonance placement (deep chest voice\,=\,calm authority; high head voice\,=\,vulnerability), and pace modulation (deliberate slowing\,=\,gravity; accelerating\,=\,anxiety). Write explicit delivery directions in the script alongside each dialogue line (\eg \texttt{[barely above a whisper, breath catching]}). Reinforce these qualities with VOX processing in post-production.
& The texture, grain, and physicality of a voice communicates character depth that no amount of descriptive narration can match; listeners form emotional connections to voice quality before they consciously process the words.
& \cite{rattigan_theatre_2002, crook_radio_2002, bernaerts_audionarratology_2021, yt-bbcscriptwriting, yt-bbcwritingforradio, yt-ryanaudiodrama} \\

\addlinespace

\textbf{Sound Design}
& Layer all four sound categories in every scene
& Treat each scene as four independent tracks, all of which must be consciously addressed: (1) \emph{VOX}---any processing applied to character voices to reflect acoustic context (\eg telephone band-pass filter, room reverb, megaphone effect, whisper treatment); (2) \emph{SFX}---discrete sound events triggered by specific actions (\eg a door opening, glass shattering, footsteps); (3) \emph{Ambience}---the continuous environmental background running under the entire scene (\eg wind, traffic, crowd murmur, rain); (4) \emph{Music}---the emotional underscore. Before producing each scene, confirm that each track has at least one intentional element assigned to it.
& Each layer contributes a distinct dimension of the story world; missing any single layer creates an incomplete aural image---a scene without ambience sounds staged and artificial; a scene without music lacks emotional direction.
& \cite{wayland2020bombs, toscan2023writingaudiodrama, crook_radio_2002, yt-bbcscriptwriting, yt-bbcwritingforradio, yt-ryanaudiodrama} \\

\addlinespace

& Foley every character--object interaction with character-specific sounds
& For every moment a character physically handles or interacts with an object, generate or record a corresponding close-proximity Foley sound. This applies to: picking up or setting down any object, footsteps (distinct per character), using a phone, handling paper, opening or closing a door, sitting in or rising from a chair, writing. Do not use generic SFX library sounds for on-scene character actions---library sounds lack intimacy and acoustic consistency with the scene. The \emph{manner} of interaction must be encoded in the sound: a character slamming a cup down versus placing it gently are two different Foley takes that communicate two different emotional states.
& The way a character physically interacts with objects expresses their emotional state through sound---the primary form of physical emotional expression in a medium without visuals.
& \cite{wayland2020bombs} \\

\addlinespace

& Establish location with ambience before any character speaks
& Begin every scene by fading in the ambience track for 2--5 seconds before dialogue starts. Select a combination of 2--3 specific environmental sounds that unambiguously signal the location (\eg hospital: HVAC hum + distant PA announcement + squeaking wheels; beach: ocean waves + seagulls + wind). Once location is established and dialogue begins, reduce ambience by 6--10\,dB so it sits beneath speech for the rest of the scene. Maintain the ambience continuously---never cut it mid-scene unless the location changes.
& Listeners need a spatial anchor before they can process dialogue effectively; beginning a scene mid-dialogue with no acoustic environment leaves the listener spatially disoriented, reducing both comprehension and immersion.
& \cite{wayland2020bombs, toscan2023writingaudiodrama, yt-writingcomics, yt-boothjunkieaudiodramasounddesign} \\

\addlinespace

& Deliberately classify each sound element as diegetic or non-diegetic
& For every sound element, ask: \emph{``Can the characters in this scene hear this?''} Diegetic [D] sounds (characters can hear them): SFX, Foley, and source music (\eg a record player in the room). Non-diegetic [ND] sounds (only the listener hears them): emotional underscore music and stylized sound design. Label each element in the script as [D] or [ND] and strictly enforce the distinction: if the underscore is non-diegetic, characters must never react to it; if a sound is diegetic, characters may respond to it in dialogue or action.
& Accidentally blurring diegetic and non-diegetic breaks the internal logic of the story world; listeners unconsciously track this consistency, and a violation produces a subliminal sense that something is wrong, eroding immersion.
& \cite{bernaerts_audionarratology_2021, crook_radio_2002, yt-ryanaudiodrama, yt-boothjunkieaudiodramasounddesign} \\

\addlinespace

& Trust the listener's imagination; avoid over-specifying every sound
& After finalizing a scene's sound design, remove each SFX element one at a time and ask: \emph{``Does removing this sound genuinely confuse the listener about what is happening?''} If the answer is no, remove it. Avoid stacking multiple simultaneous SFX for a single narrative moment. Leave deliberate sonic gaps for the listener's imagination to fill in. Guiding rule: not every physical action needs a corresponding SFX---a character walking to a window may need only a subtle ambience shift, not explicit footstep Foley.
& Audio drama uniquely activates the listener's imagination in a way no visual medium can; over-specifying every sound replaces imaginative engagement with passive reception, reducing immersion rather than increasing it.
& \cite{crook_radio_2002, rattigan_theatre_2002, wayland2020bombs, yt-cactusaudiodrama} \\

\addlinespace

\textbf{Music}
& Underscore each scene's emotional tone with scene-specific music
& Before selecting or composing music, write a single-word emotional descriptor for each scene (\eg ``dread,'' ``tenderness,'' ``urgency,'' ``relief''). Choose music that precisely matches this descriptor---not generic background music. Introduce the music cue at scene start or at the moment of an emotional shift mid-scene. Set music level $\approx$\,15--18\,dB below dialogue. Attenuate music in the 250--3000\,Hz range to prevent frequency masking of speech. Change the music cue whenever the scene's emotional tone shifts substantially.
& Music operates on listeners' emotions faster and more reliably than dialogue; the right underscore ensures listeners feel a scene's emotional intent before they consciously process the words, making emotional engagement more immediate.
& \cite{wayland2020bombs, toscan2023writingaudiodrama, de_fossard_writing_2005, crook_radio_2002, yt-bbcscriptwriting, yt-bbcwritingforradio} \\

\addlinespace

& Use music to establish acoustic presence in interior scenes with no natural ambience
& For interior scenes in acoustically neutral settings (\eg an empty office, a bare room) where no recognizable environmental sound is available, use a sustained, low-energy musical pad or synthesized electronic texture as a substitute ambience. Place it at ambience level ($\approx$\,$-$45 to $-$50\,LUFS). Match the tone to the scene's emotional register: a warm, slow pad for intimacy; a cold, high-frequency drone for unease. If any faint location-specific sound exists (\eg HVAC hum, distant traffic), layer it with the musical pad.
& Complete acoustic silence beneath dialogue reads as a production error rather than an artistic choice; it creates listener discomfort and collapses the sense of spatial presence within the scene.
& \cite{toscan2023writingaudiodrama, wayland2020bombs, yt-cactusaudiodrama} \\

\addlinespace

& Establish recurring musical motifs for each major character and the show itself
& Compose or select a distinctive 5--10 second melodic or rhythmic motif for each major character. Play it at underscore level whenever the character is introduced or is the scene's emotional focus. Create a 20--30 second signature theme tune for the show's opening. Use a 2--3 second transition sting (a short, resolving musical phrase) as the standard sonic device between scenes. Repeat each motif consistently and never substitute it---listeners must form an unconscious association between motif and character.
& Recurring motifs create emotional shorthand: after a few scenes, hearing a character's motif alone triggers the listener's full emotional association with that character, deepening narrative investment without any additional dialogue.
& \cite{de_fossard_writing_2005, crook_radio_2002} \\

\addlinespace

\textbf{Scene Transitions \& Acoustic Space}
& Communicate every scene change through sound, not dialogue
& Use two techniques together: (1) \emph{Contrasting ambiences}---fade out Scene A's ambience, then fade in Scene B's distinctly different ambience (\eg busy caf\'{e} $\rightarrow$ quiet forest); (2) \emph{Carrier sounds}---use a sound that physically performs the transition (footsteps walking out, a door closing, a car driving away, a phone hanging up). Insert a brief 1--2 second silence or a transition sting between the two ambiences to mark the cut. Never write dialogue that telegraphs a scene change (\eg \emph{``Well, I should be going now''}).
& Dialogue-telegraphed transitions are clunky and unnatural; sound-based transitions are seamless and preserve dramatic momentum without requiring the listener to consciously acknowledge a location change.
& \cite{wayland2020bombs, toscan2023writingaudiodrama, crook_radio_2002, rattigan_theatre_2002, yt-writingcomics, yt-cactusaudiodrama} \\

\addlinespace

& Signal temporal jumps through a consistent, distinctive sound device
& Design a single unique temporal-shift audio signature and use it \emph{every time} the narrative moves forward or backward in time. Options include: a distinct musical sting with a long reverb tail; an abrupt cut to silence followed by a different acoustic texture; a pitch-shifted or time-stretched sound sweep; or a shift from stereo to mono for scenes set in the past. For flashback sequences, apply a consistent audio treatment throughout (\eg warmer EQ, more compression, longer reverb decay). Establish this device in the first episode and never deviate from it.
& Audio drama has no visual chapter breaks or title cards; without an explicit audio cue, a temporal jump leaves listeners confused about where they are in the narrative timeline---confusion that compounds if left unresolved.
& \cite{bernaerts_audionarratology_2021} \\

\addlinespace

& Use acoustic perspective to place each character physically in the scene space
& Apply consistent acoustic positioning: \emph{Close/foreground} (arm's reach): dry voice, no reverb, full frequency range. \emph{Across the room}: add medium room reverb, slightly reduce presence frequencies (2--5\,kHz), lower volume 4--6\,dB. \emph{Through a wall}: add heavy reverb, high-frequency rolloff above 4\,kHz, lower volume 8--12\,dB. \emph{On a telephone}: apply a band-pass filter (300--3400\,Hz), add mild distortion, compress dynamic range. Maintain each character's assigned acoustic position consistently within a scene; shift it only if they physically move in the narrative.
& Without visual cues, all characters default to sounding equally close to the listener; acoustic positioning is the only tool for creating spatial realism and preventing the scene from sounding like a flat, stageless recording.
& \cite{crook_radio_2002, bernaerts_audionarratology_2021} \\

\addlinespace

\textbf{Acoustic Focalization \& Listener Agency}
& Choose whose acoustic point of view the listener inhabits in each scene
& At the top of each scene in the script, write a \emph{focalization note}: \emph{``Listener hears through [Character]'s perspective.''} Implement this by: (a) including only sounds that character could physically hear from their position in the scene; (b) using that character's acoustic location as the spatial reference point for all other sounds; (c) withholding sounds from events happening elsewhere to create deliberate information asymmetry. For dramatic irony, momentarily shift focalization to include a sound the focal character cannot hear (\eg the quiet sound of a door opening behind them while they speak).
& Acoustic focalization controls information flow, builds suspense, and enables dramatic irony---the audio equivalent of film's camera POV. Used deliberately, it creates a uniquely intimate listener--character bond.
& \cite{bernaerts_audionarratology_2021} \\

\addlinespace

& Use deliberate silence as an active dramatic tool
& At moments of high dramatic weight---a revelation, a key decision, an unexpected twist---simultaneously cut all ambience, music, and SFX tracks, inserting 2--4 seconds of near-complete silence. Mark this explicitly in the script as \texttt{[SILENCE: 3\,sec]}. Build toward the silence by gradually removing sound elements beforehand (remove music first, then SFX, then ambience) to amplify anticipation. Distinguish planned silence from a production gap by ensuring it is precisely timed and followed by a deliberate, impactful sound event.
& The sudden withdrawal of all sound forces the listener's full attention, creates a physical sense of anticipation, and makes the subsequent sound---a single word, a gunshot, one musical note---maximally impactful.
& \cite{crook_radio_2002, rattigan_theatre_2002, bernaerts_audionarratology_2021} \\

\midrule

\multicolumn{5}{l}{\textbf{DG3: Ensure production-quality audio mixing}} \\
\midrule

\textbf{Production Consistency}
& Maintain acoustic and character consistency throughout the entire production
& Create and maintain a \emph{production bible} documenting for each character: voice actor, microphone model and position, recording room, and EQ/compression settings. Document for each recurring location: the ambience file name and version, and reverb settings (room size, pre-delay, decay). Reuse these exact settings every time a character or location reappears. Never substitute a different voice actor mid-production; if a re-recording is necessary, precisely match the original session's acoustic conditions.
& Acoustic inconsistency---a character sounding different between scenes, or the same location sounding different across episodes---immediately breaks immersion; listeners detect these mismatches unconsciously even when they cannot articulate what is wrong.
& \cite{de_fossard_writing_2005, crook_radio_2002, yt-boothjunkieaudiodramasounddesign} \\

\addlinespace

\textbf{Audio Mixing}
& Apply a consistent audio mixing hierarchy so that dialogue always remains intelligible
& Target these integrated loudness levels (LUFS) in the final mix: \emph{Dialogue/Voice}: $-$24\,LUFS (reference level; must never be masked by other elements). \emph{Music underscore}: $-$30\,LUFS ($\approx$\,6\,dB below voice). \emph{SFX}: $-$33 to $-$39\,LUFS (adjusted by narrative importance). \emph{Ambience}: $-$45 to $-$60\,LUFS (background only). Additionally, apply a high-pass filter to music at 200--300\,Hz and a low-pass at 8--10\,kHz to prevent music frequencies from masking speech. Verify the final mix on at least two playback systems (headphones and loudspeakers).
& If music or SFX mask dialogue at any moment, the primary content of the drama---what characters say---is irreversibly lost for the listener; the mixing hierarchy ensures each sound layer serves its role without interfering with the others.
& \cite{de_fossard_writing_2005, crook_radio_2002, yt-cactusaudiodrama} \\

\midrule

\multicolumn{5}{l}{\textbf{DG4: Support iterative refinement across script and audio}} \\
\midrule

\textbf{Production Workflow}
& Write the script as a complete production blueprint, not just a dialogue document
& Use a standardized script format throughout: character names in \textsc{Small Caps} flush left, dialogue below, and all sound cues in [SQUARE BRACKETS] at the exact script position where they occur. Each sound cue annotation must specify: (a) type (SFX / AMBIENCE / MUSIC / VOX), (b) a precise description---not ``car sound'' but \emph{``[SFX: heavy diesel engine idling, then cutting out abruptly]''}, (c) its approximate duration or whether it is continuous, (d) its emotional function (\eg \emph{``[MUSIC: low tension underscore, establishes dread---continues under dialogue]''}). Include a Character Voice Profile page and a Locations \& Ambience reference page at the front of the script.
& Ambiguous scripts require verbal clarification between writers, directors, and sound designers, introducing inconsistencies and rework; a fully annotated script serves as a complete production specification that enables direct translation from script to audio.
& \cite{de_fossard_writing_2005, yt-cactusaudiodrama, yt-boothjunkieaudiodramasounddesign} \\

\end{longtable}

\twocolumn

\begin{table}
\centering
\small
\begin{tabularx}{\linewidth}{lX}
\toprule
\textbf{Dimension} & \textbf{What is scored} \\
\midrule
Dialogue efficiency & Lines are short ($\leq$17 words), natural, one thought per turn \\
Narration discipline & Narrator used only when information cannot be conveyed using other sounds \\
Exposition handling & Backstory scattered in dialogue, never front-loaded \\
Narrative clarity & Zero-background listener can follow; each line responds to the prior one \\
Scene propulsion & Every scene opens mid-action and ends on unresolved tension \\
SFX specificity & Prompts name material and manner (\eg ``leather soles scraping bark'') \\
SFX discipline & SFX are spaced, not stacked; each is necessary \\
Sound design coverage & Every scene has at least one SFX or music cue \\
Music emotional precision & Cue prompts name the emotion; placement aligns with tonal shifts \\
Adaptation fidelity & All key events preserved; internal thought externalized \\
Cross-modal redundancy & No text restates what a co-located SFX already conveys \\
\bottomrule
\end{tabularx}
\caption{11 rubric dimensions used by the LLM judge to score script candidates. Each dimension is rated on a 1--5 scale.}
\label{tab:rubric}
\end{table}

\begin{table*}
  \footnotesize
  \renewcommand{\arraystretch}{1.35}
  \begin{tabularx}{\textwidth}{@{} c c c l r c >{\raggedright\arraybackslash}X >{\raggedright\arraybackslash}p{5cm} @{}}
  \toprule
  \textbf{ID} & \textbf{Age} & \textbf{Gender} & \textbf{Country} & \textbf{Experience} & \textbf{Expertise*} & \textbf{Professional Background} & \textbf{Frequently used Generative AI Tools} \\
  \midrule
  P1 & 31 & M & Morocco       & 12 years & 3 & Audio series, film production, and dubbing                                    & ElevenLabs \\
  P2 & 49 & F & USA           &  1 year & 2 & Audio editing, noise removal, and mixing                                      & ElevenLabs, Descript \\
  P3 & 47 & M & Greece        &  8 years & 3 & Music composition and sound design (children's, sci-fi)                        & ElevenLabs \\
  P4 & 56 & M & UK            &  3 years & 3 & Audiobook narration and production (children's)                                & ElevenLabs \\
  P5 & 39 & M & N.\ Macedonia &  3 years & 3 & Sound design, music integration, and voiceover audiobooks                      & ElevenLabs, Cartesia, Stable Audio \\
  P6 & 20 & M & Costa Rica    &  1 year & 2 & Mixing and sound design (thrillers, mystery)                                   & ElevenLabs, Suno \\
  P7 & 29 & M & Montenegro    &  4 years & 3 & Audio production and voiceover                                                & ElevenLabs \\
  P8 & 44 & M & Philippines   & 17 years & 3 & Directing, sound design for radio drama, podcasts, and commercials (sci-fi, action, crime, horror, fantasy) & ElevenLabs, Suno, Udio, Hume, Gemini \\
  \bottomrule
  \end{tabularx}
  \caption{User study participants (P1--P8). All participants were recruited from Upwork with 6.1 years of audio editing experience ($\sigma = 5.8$) and all have created or edited audio dramas or immersive audio content before. *Self-reported audio editing expertise (1\,=\,Beginner, 2\,=\,Intermediate, 3\,=\,Advanced).}
  \Description{Table listing the 8 user study participants with their age, gender, country, years of audio drama experience, self-reported expertise level, professional background, and generative AI tools used in their existing workflows.}
  \label{tab:participants}
  \end{table*}

\begin{table}
\footnotesize
\renewcommand{\arraystretch}{1.35}
\begin{tabularx}{\linewidth}{@{} c >{\raggedright\arraybackslash}p{2.5cm} >{\raggedright\arraybackslash}X @{}}
\toprule
\textbf{ID} & \textbf{Background} & \textbf{Relevant Projects} \\
\midrule
N1 & Escape room designer     & Digital and physical escape rooms, gamified e-books, interactive learning content \\
N2 & Tabletop RPG enthusiast  & DnD campaigns, narrative RPG scenarios, interactive fiction \\
N3 & Game designer & Game production, dialogue scripting, level and scene design \\
\bottomrule
\end{tabularx}
\caption{Informal exploratory study participants (N1--N3). None had prior audio production experience.}
\Description{Table listing the 3 informal exploratory study participants with their background and relevant project domains.}
\label{tab:participants-exploratory}
\end{table}

\begin{table}
\resizebox{\linewidth}{!}{%
\begin{tabular}{llll}
\toprule
\textbf{Measure} & \textbf{Existing Practice} & \textbf{\system{}} & \textbf{$p$} \\
\midrule
\textbf{NASA-TLX} &  &  &  \\
Mental Demand & $\mu = 4.25$, $\sigma = 0.89$ & $\mu = 2.00$, $\sigma = 1.07$ & $.0078$ $**$ \\
Temporal Demand & $\mu = 3.88$, $\sigma = 0.35$ & $\mu = 2.25$, $\sigma = 0.89$ & $.0156$ $*$ \\
Performance & $\mu = 4.38$, $\sigma = 1.06$ & $\mu = 3.88$, $\sigma = 0.83$ & $.3438$ \\
Effort & $\mu = 4.62$, $\sigma = 0.74$ & $\mu = 1.75$, $\sigma = 0.89$ & $.0078$ $**$ \\
Frustration & $\mu = 3.00$, $\sigma = 1.31$ & $\mu = 1.88$, $\sigma = 0.83$ & $.1250$ \\
&  &  &  \\
\textbf{Creativity Support Index} &  &  &  \\
Exploration & $\mu = 2.62$, $\sigma = 0.92$ & $\mu = 4.12$, $\sigma = 1.13$ & $.0391$ $*$ \\
Engagement & $\mu = 4.62$, $\sigma = 1.06$ & $\mu = 4.38$, $\sigma = 0.92$ & $1.000$ \\
Worth the Effort & $\mu = 4.38$, $\sigma = 0.92$ & $\mu = 4.88$, $\sigma = 0.35$ & $.3750$ \\
Transparency & $\mu = 3.00$, $\sigma = 1.31$ & $\mu = 3.88$, $\sigma = 1.36$ & $.0625$ \\
Expressiveness & $\mu = 4.62$, $\sigma = 0.52$ & $\mu = 4.00$, $\sigma = 1.07$ & $.2500$ \\
&  &  &  \\
\textbf{System Usability Scale} &  &  &  \\
Frequent Use & $\mu = 4.00$, $\sigma = 0.76$ & $\mu = 4.62$, $\sigma = 0.74$ & $.2812$ \\
Easy to use & $\mu = 2.88$, $\sigma = 1.25$ & $\mu = 4.62$, $\sigma = 0.52$ & $.0312$ $*$ \\
Quick to learn & $\mu = 1.62$, $\sigma = 0.74$ & $\mu = 4.38$, $\sigma = 0.74$ & $.0078$ $**$ \\
Confident Use & $\mu = 4.50$, $\sigma = 0.76$ & $\mu = 4.38$, $\sigma = 1.06$ & $1.000$ \\
Lot to learn & $\mu = 4.25$, $\sigma = 1.49$ & $\mu = 2.00$, $\sigma = 1.07$ & $.0156$ $*$ \\
\bottomrule
\end{tabular}%
}
\caption{Within-subject comparison of existing practice and \system{} ($n=8$). We report means ($\mu$) and sample standard deviations ($\sigma$) per condition, and $p$-values from two-tailed paired Wilcoxon Signed Rank tests. $^{*}p < .05$, $^{**}p < .01$.}
\label{tab:ratings_significances}
\end{table}

\end{document}